\documentclass[aps,pre,preprint,superscriptaddress]{revtex4-2}
\usepackage[utf8]{inputenc}
\usepackage[T1]{fontenc}
\usepackage{graphicx}
\usepackage{xcolor}
\usepackage{amssymb,amsfonts,amsmath}
\usepackage{amsmath}
\usepackage{tikz}
\usetikzlibrary{arrows.meta}

\begin{document}

\title{Diversification under Global Competition: A Cutoff-Induced Localization Mechanism}

\author{David A. Kessler}
\affiliation{Department of Physics, Bar-Ilan University, Ramat Gan 5290002, Israel}
\author{Nadav M. Shnerb}
\affiliation{Department of Physics, Bar-Ilan University, Ramat Gan 5290002, Israel}

\begin{abstract}
Niche differentiation promotes biological diversity but does
not necessarily eliminate competition for shared limiting
resources. What determines the extent of diversity when
evolutionary diversification takes place under such constraints?
We investigate this question using a simple, analytically
tractable model combining mutations in continuous trait space,
local competition, and global competition for shared resources.
We show that, without an establishment threshold, the population
spreads indefinitely in trait space even as its total abundance
approaches a finite limit. In contrast, an arbitrarily small
establishment threshold allows a stable localized distribution.
Together with mortality induced by global
competition, the threshold creates bistability, and global
feedback drives the system exponentially close to the Maxwell
point, where an isolated invasion front is stationary.
The resulting soliton has height
$H\propto\varepsilon^{2/3}$ and width
$W\propto\varepsilon^{-2/3}$.
Simulations with discrete individuals reveal
an archipelago of fluctuating clusters whose spatial extent
eventually reaches statistical saturation, unlike the continuous
deterministic soliton. Nevertheless, an effective cutoff
$\varepsilon_{\mathrm{eff}}\simeq0.8/(N\sqrt{\mu})$
reproduces the observed scaling of the population mass deficit.
Remarkably, both occupied trait-site richness and effective
species richness exhibit approximately the same scaling as
the deterministic soliton width. Thus, deterministic cutoff
theory captures global diversity scaling despite failing
to reproduce the community's spatial organization.
\end{abstract}

\maketitle
\clearpage
\section{Introduction}
\label{sec:introduction}

The coexistence of what Darwin called “endless forms most beautiful and most wonderful” is a source of wonder, but also poses a scientific puzzle: how can such an extraordinary diversity be sustained in a world of limited resources? One of the central answers offered by evolutionary
ecology is niche differentiation: species with different
traits may specialize in exploiting different resources,
thereby reducing competition among themselves
\citep{dieckmann1999origin,scheffer2006self}. Yet in many
ecological systems, niche differentiation does not eliminate
competition for shared resources. Two tree species
specializing in different nutrients  still compete for
water, sunlight, and space; in general, species occupying
distinct niches may remain dependent on common limiting
resources. What, then, determines the species richness that
can be sustained under such conditions? Under what
circumstances can evolutionary diversification continue
despite competition for shared resources?

Ecological limits to diversification need not arise from a shortage of available niches. Storch and Okie argued that, although the addition of species may increase total community abundance by allowing more complete utilization of resources, this increase is ultimately constrained by finite resource and energy availability. Consequently, as species richness grows, individual populations tend to become smaller and more vulnerable to extinction, potentially limiting diversity even when niche space remains unsaturated \citep{storch2019carrying}. A complementary perspective was provided by Tilman, who showed that the establishment of new species depends on the resources left available by resident species within a habitat, as well as on the stochastic survival of initially rare colonists \citep{tilman2004niche}.

Theoretical studies have also shown how trait-dependent competition can lead to differentiation and clustering in niche space \citep{scheffer2006self,pigolotti2007species}. Yet the emergence of distinct niches does not, by itself, determine how far diversification can proceed when populations occupying different niches remain constrained by shared resources. This raises a more specific question: how do global competition for shared resources and the ability of initially small populations to establish themselves in previously unoccupied niches jointly determine the extent of diversification in trait space?

Here we introduce a simple, analytically tractable model
of evolutionary diversification. Mutations allow a
population to spread along a continuous axis of ecological
traits; local competition limits the density of individuals
with similar traits, while global competition for shared
resources constrains the total population size. We examine
how these ingredients determine the extent of ecological
diversity, represented in the model by the width of the
trait distribution. In the simplest version of the dynamics, the population
continues to spread indefinitely in trait space even though
its total mass approaches a finite limit. Consequently,
an ever wider range of niches is occupied at progressively
lower average density.

The unbounded spreading predicted by the deterministic continuum model relies on the assumption that reproduction remains possible at arbitrarily low local densities. In real populations, reproduction at low densities may be suppressed by an ecological Allee effect \citep{keitt2001allee}; moreover, even in the absence of such an effect, demographic discreteness limits the validity of the continuum description and  generates an effective cutoff \citep{brunet1997shift,perthame2010survival}. We therefore begin by examining the deterministic consequences of a sharp establishment threshold, below which reproduction is suppressed, and subsequently investigate what changes when demographic fluctuations are treated explicitly.
We show that even an arbitrarily small threshold
qualitatively changes the dynamics, allowing a stable
localized trait distribution. The analytical tractability
of the model enables us to construct the family of
stationary localized solutions, determine their conditions
of existence, and examine which branch is dynamically
selected. Physically, the establishment cutoff,
\emph{together with} the mortality imposed by global
competition, creates bistability between empty and
populated states. Global competition then provides feedback
that nearly balances the invasion of empty niches against
the retreat of populated ones, driving the system
exceptionally close to the Maxwell point at which an
isolated front is stationary. We show that the distance
from this point is exponentially small in the inverse
global competition strength (with a cutoff-dependent
scale), whereas the width of the localized profile
diverges logarithmically as the Maxwell point is approached.
These results lead to nontrivial scaling laws: as the
threshold $\varepsilon$ tends to zero, the profile height
scales  as $H\propto\varepsilon^{2/3}$, while its width scales
as $W\propto\varepsilon^{-2/3}$.

Finally, we return to the demographic interpretation of the
establishment threshold and ask how faithfully a deterministic
cutoff can describe a population composed of discrete individuals.
Previous studies have shown that demographic fluctuations can
generate clusters in trait space even when the corresponding
deterministic dynamics does not predict  amplitude oscillations
\citep{rogers2012demographic,ganan2018front}.
We make a quantitative comparison between the individual-based
model and the deterministic soliton, whose properties have been
characterized analytically. We find that whereas the deterministic
cutoff produces a continuous, localized profile, the stochastic
dynamics generates an  ever-changing archipelago of separated clusters whose
spatial extent eventually reaches statistical saturation. Despite this fundamental difference in spatial organization, the analytical soliton prediction, with the cutoff identified as an effective discreteness scale, approximately captures the scaling of the population mass deficit in the individual-based model. This deficit, $g=1-\kappa\langle M\rangle$, measures the fractional shortfall of the mean total population mass $\langle M\rangle$ from its limiting value $1/\kappa$ in the absence of a cutoff. Moreover, the number of occupied trait niches and the
effective species richness exhibit approximately the same scaling
as the deterministic soliton width, although the spatial extent
of the stochastic population is substantially larger.
These results demonstrate that a simple deterministic description
can capture the scaling laws of global ecological properties
even when it fails to reproduce the spatial organization
of the community.

\section{A solvable one-dimensional continuum model of lineage localization}
\label{sec:continuum}

In this section, we introduce a minimal continuum model of evolution in trait space. We assume that $x$ represents a phenotypic trait and that $\phi(x,t)$ denotes the population density along this trait axis at time $t$. During reproduction, offspring inherit their parents' traits with small random modifications, so that, in the continuum limit, mutations are represented by diffusion in trait space with diffusion coefficient $D$. We assume that competition has two components: local competition between individuals with similar traits, represented by a quadratic density-dependent term, and global competition for a shared limiting resource, which persists even as organisms diversify into distinct ecological niches. The strength of global competition is proportional to the total population, $M(t)=\int\phi(x,t)\,dx$. Finally, we introduce a local establishment threshold $\varepsilon$, below which reproduction is suppressed while competition-induced mortality remains operative. This threshold admits two distinct interpretations. It may represent a genuine ecological constraint, such as an idealized Allee effect, in which small local populations cannot reproduce effectively. Alternatively, it may serve as a deterministic approximation to demographic discreteness, following the cutoff approach of Brunet and Derrida~\cite{brunet1997shift}. These ingredients lead to a nonlinear reaction--diffusion equation, a nonlocal form of the  cut-off Fisher-KPP equation~\cite{berestycki2009nonlocal} which we use to investigate how establishment thresholds and global competition for shared resources jointly regulate the spread of an evolving population in trait space.

Under these assumptions, the dynamics of the population density in trait space is governed by
\begin{equation}
\frac{\partial\phi(x,t)}{\partial t}
=D\frac{\partial^2\phi(x,t)}{\partial x^2}
+\phi(x,t)\,\Theta\bigl(\phi(x,t)-\varepsilon\bigr)
-\phi^2(x,t)
-\kappa M(t)\phi(x,t),
\label{eq:continuum}
\end{equation}
where
\begin{equation}
M(t)=\int_{-\infty}^{\infty}\phi(x,t)\,dx.
\label{eq:mass}
\end{equation}
Here, $\Theta$ is the Heaviside step function, and $\kappa$ controls the strength of global competition. We have chosen units in which both the reproduction rate and the coefficient of local competition are unity. Reproduction therefore occurs only where the population density exceeds the threshold $\varepsilon$, whereas mortality due to local and global competition operates at all densities. Crucially, the global competition term depends on the integral over the entire trait space, so that populations occupying widely separated regions of this space continue to compete for the same shared resource.

We wish to determine whether the dynamics can support a stationary state in which the population remains localized within an effectively bounded region of trait space, despite the continuous spreading induced by mutations. Such a state does not imply that reproduction and mutation have ceased, but rather that a dynamical balance has been established in which the trait distribution remains constant in time. We therefore seek a stationary profile $\phi(x,t)=\phi(x)$ with finite total mass, satisfying $\phi(x)\geq0$ and $\phi(x)\to0$ as $|x|\to\infty$. Since the model is translationally invariant, we may center the profile at $x=0$ and seek a symmetric solution. Importantly, the strength of global competition is determined by the total population in addition to $\kappa$. Defining $r\equiv\kappa M$, we first seek localized stationary solutions for a given $r$, and then impose the self-consistency condition
\begin{equation}
r=\kappa\int_{-\infty}^{\infty}\phi_r(x)\,dx,
\label{eq:selfconsistency}
\end{equation}
where $\phi_r(x)$ denotes the stationary profile obtained for that value of $r$.

In the stationary state, the equation of motion  then becomes
\begin{equation}
D\frac{d^2\phi}{dx^2}
+\phi\Theta(\phi-\varepsilon)
-\phi^2-r\phi=0.
\label{eq:stationary}
\end{equation}

To find localized solutions, we employ the standard technique of a mechanical analogy
in which the trait coordinate $x$ plays the role of time,
while $\phi(x)$ represents the position of a particle moving, with no friction,
in an effective potential $U(\phi)$. The stationary equation
can then be written as
\begin{equation}
D\frac{d^2\phi}{dx^2}
=-\frac{dU}{d\phi}.
\label{eq:mechanical}
\end{equation}

The effective potential is obtained by integrating
Eq.~\eqref{eq:mechanical}. The integration constants are
fixed by choosing $U(0)=0$ and requiring the potential
to be continuous at $\phi=\varepsilon$, where the
reproduction term switches on. This gives
\begin{equation}
U(\phi)=
\begin{cases}
-\dfrac{r}{2}\phi^2-\dfrac{1}{3}\phi^3,
& 0\leq\phi<\varepsilon,\\[8pt]
\dfrac{1-r}{2}\phi^2-\dfrac{1}{3}\phi^3
-\dfrac{\varepsilon^2}{2},
& \phi\geq\varepsilon.
\end{cases}
\label{eq:potential}
\end{equation}

The mechanical analogy provides a simple way to identify
localized stationary solutions. Since we seek a profile
for which both $\phi$ and its derivative vanish as
$x\to\pm\infty$, and since $U(0)=0$, conservation
of mechanical energy requires
\begin{equation}
\frac{D}{2}\left(\frac{d\phi}{dx}\right)^2+U(\phi)=0.
\label{eq:energy}
\end{equation}

A localized solution corresponds to a homoclinic orbit:
the particle starts at $\phi=0$ as $x\to-\infty$,
moves towards positive values of $\phi$, reaches a
turning point $\phi=A$, and returns to $\phi=0$ as
$x\to+\infty$. Choosing the turning point at $x=0$,
the corresponding population profile satisfies
\begin{equation}
\phi(-\infty)=\phi(+\infty)=0,\qquad
\phi(0)=A,\qquad \phi'(0)=0.
\end{equation}
The turning point, at which the kinetic energy vanishes, thus satisfies
\begin{equation}
U(A)=0.
\label{eq:turning}
\end{equation}

\begin{figure}[!htb]
    \centering
    \includegraphics[width=0.62\columnwidth]{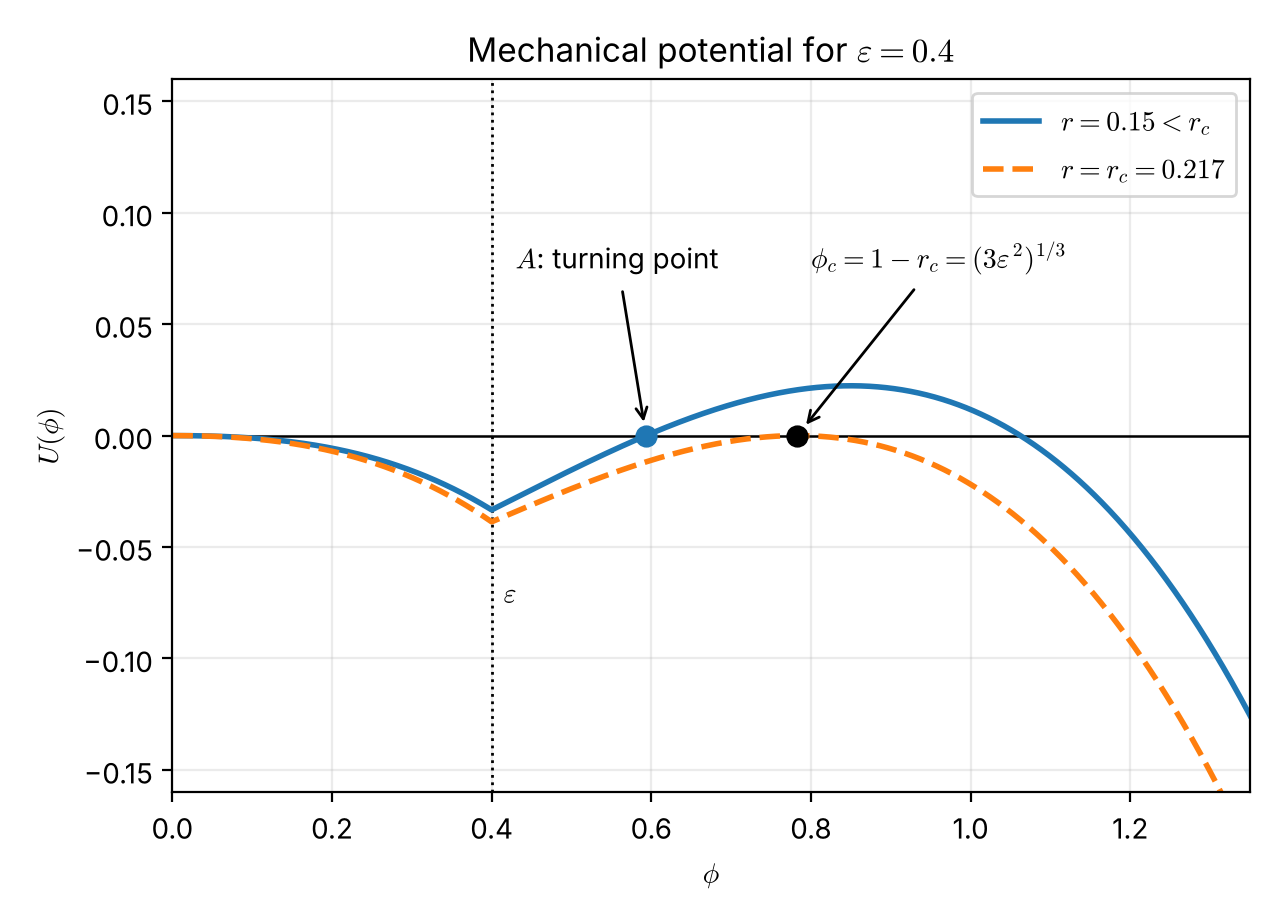}
    \caption{\textbf{Mechanical potential of the one-dimensional continuum model.} A representative example for $\varepsilon=0.4$. For $r<r_c$ (solid), the zero-energy orbit leaves $\phi=0$, reaches the first positive zero $A$ of $U$, and returns, giving a localized stationary pulse. At $r=r_c$ (dashed), the right-hand maximum touches $U=0$ at $\phi_c=(3\varepsilon^2)^{1/3}$; the turning point becomes marginal and the pulse width diverges. For $r>r_c$, the maximum lies below zero and no zero-energy localized pulse exists.}
    \label{fig:continuum_potential}
\end{figure}

 Eq.~\eqref{eq:energy} implies
\begin{equation}
\left(\frac{d\phi}{dx}\right)^2=-\frac{2U(\phi)}{D}.
\end{equation}
Thus, a localized solution requires the potential
to be negative between the origin and the turning point,
\begin{equation}
U(\phi)<0,\qquad 0<\phi<A.
\label{eq:localization_condition}
\end{equation}
As $\phi$ increases from 0, the potential initially decreases
till $\phi=\varepsilon$, then starts rising and eventually reaches a maximum
beyond which it decreases (Fig.~\ref{fig:continuum_potential}).
Thus the maximum value of $U$ has to be positive for there to be a turning point $A$. The local maximum of the potential
occurs at $\phi_m=1-r$, where
\begin{equation}
U(\phi_m)=\frac{(1-r)^3}{6}-\frac{\varepsilon^2}{2}.
\end{equation}
Requiring $U(\phi_m)>0$ yields
\begin{equation}
0<r< r_c(\varepsilon)
\equiv 1-3^{1/3}\varepsilon^{2/3},
\label{eq:rc}
\end{equation}
provided the cutoff is sufficiently small, $\varepsilon < 1/\sqrt{3}$, so that $r_c>0$.

For every $0<r<r_c$, we have, therefore, a localized
stationary solution $\phi_r(x)$. However, in the
original model, $r$ is not an independent parameter
but must satisfy the self-consistency condition (Eq. \ref{eq:selfconsistency})
$r=\kappa M$. Thus, each soliton in this family
can be associated with a global competition strength
for which it is a stationary solution:
\begin{equation}
\kappa(r)=\frac{r}{M(r)},\qquad
M(r)=\int_{-\infty}^{\infty}\phi_r(x)\,dx.
\label{eq:kappa_of_r}
\end{equation}

\begin{figure}[!htb]
    \centering
    \includegraphics[width=0.7\columnwidth]{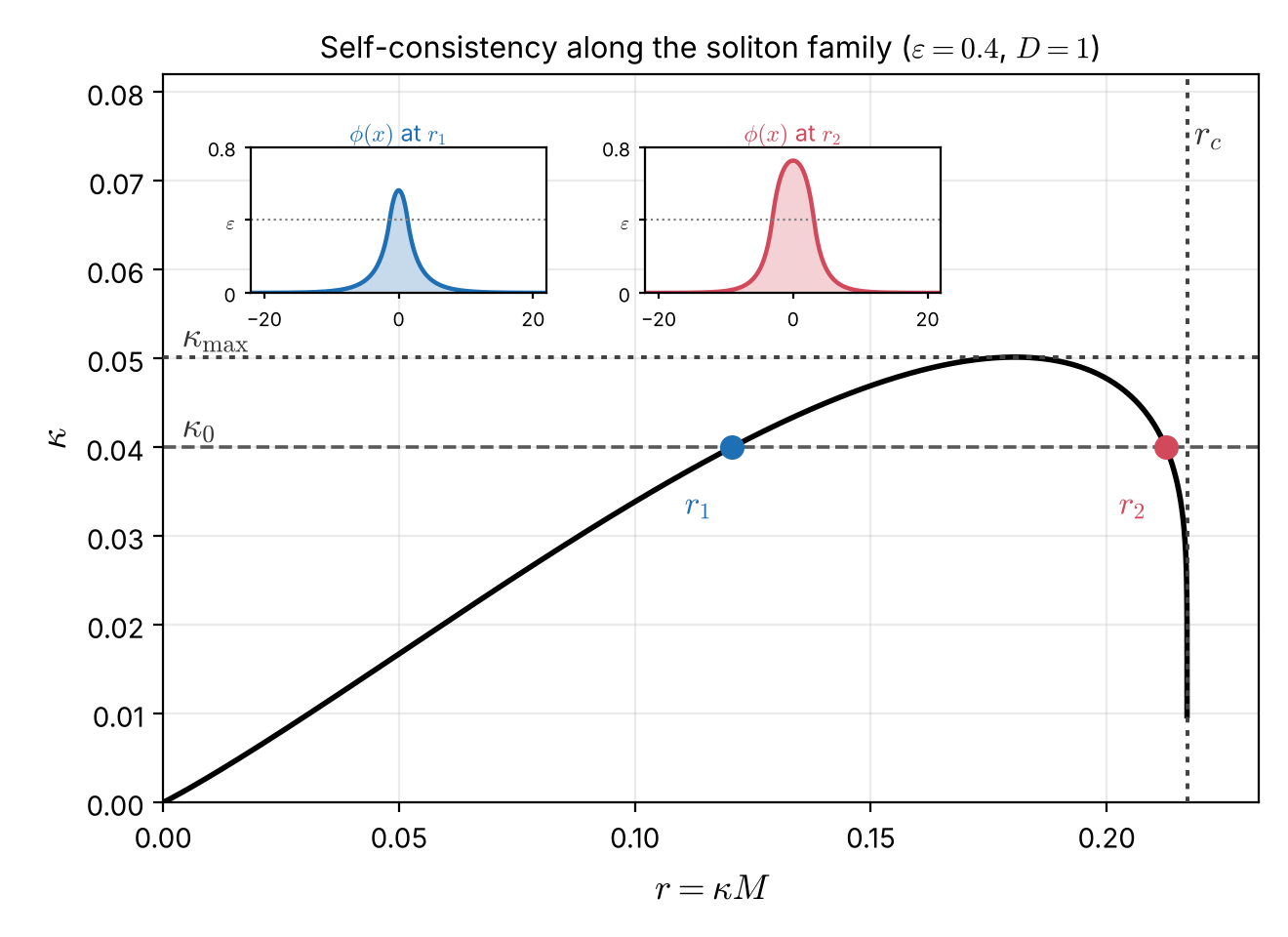}
    \caption{\textbf{Self-consistency along the family of localized stationary solutions.} For $\varepsilon=0.4$ and $D=1$, each $0<r<r_c$ admits a localized solution $\phi_r(x)$ with total mass $M(r)$. The solid curve shows the corresponding global competition strength, $\kappa(r)=r/M(r)$. For the chosen value $\kappa_0=0.04$ (horizontal dashed line), two intersections are obtained, at $r_1\simeq0.12$ and $r_2\simeq0.21$, representing two distinct stationary solutions of the same model. The maximum $\kappa_{\max}\simeq0.05$ and the critical value $r_c\simeq0.22$ are indicated by dotted lines. The insets show the corresponding density profiles, with dotted horizontal lines marking the reproduction cutoff $\varepsilon$.  As $r\to r_c^-$, the mass diverges logarithmically and $\kappa(r)$ approaches zero; this limiting behavior occurs on a scale too narrow to be resolved in the main plot.}
    \label{fig:kappa_selfconsistency}
\end{figure}

Figure~\ref{fig:kappa_selfconsistency} shows $\kappa(r)$ for $\varepsilon=0.4$
and $D=1$. The function is nonmonotonic and exhibits
a maximum. Consequently, a given global competition
strength may admit either zero or two distinct stationary solutions.
For example, the horizontal line corresponding to
$\kappa=0.04$ intersects the curve at two points,
$r_1$ and $r_2$, representing solitons with different
heights and widths, as illustrated by the profiles
shown in the insets. The existence of these stationary
solutions does not, by itself, establish their
dynamical stability.

To understand the behavior of the localized solutions,
we first consider the limit in which the reproduction
cutoff $\varepsilon$ approaches zero. In the absence
of this cutoff, our model reduces to a Fisher-type
reaction--diffusion equation with global competition
between individuals, regardless of their separation
in trait space. While global competition limits the
total population, it cannot prevent its spreading
in trait space. In particular, as we demonstrate
in Appendix~\ref{app:no_cutoff}, even in the presence of infinite-range
interactions, the population distribution continues
to spread indefinitely when $\varepsilon=0$.
Since the total population remains bounded, this
spreading is accompanied by a progressive decrease
in population density. Thus, the reproduction cutoff
$\varepsilon$ provides the mechanism that allows
a localized stationary profile to exist, and we
expect its width to diverge as $\varepsilon\to0$.
We therefore focus on the broad branch of stationary
solutions, which turns out to be dynamically stable, whereas
the narrow branch is unstable, as discussed in
Appendix~\ref{app:stability}.

 From the mechanical
analogy, we know that this divergence occurs as $r$
approaches its critical value $r_c$ from below.
In this limit, the particle spends an increasingly
long time near the maximum of the effective potential,
producing a broad, nearly flat population profile
whose height approaches the position of that maximum.

Using the critical value $r_c$ obtained in Eq.~\eqref{eq:rc},
the soliton height satisfies
\begin{equation}
H\equiv\max_x\phi(x)
\simeq 1-r_c
=(3\varepsilon^2)^{1/3}.
\end{equation}

On the other hand, the self-consistency condition
requires
\begin{equation}
M=\frac{r}{\kappa}
\simeq\frac{r_c}{\kappa}
\longrightarrow\frac{1}{\kappa}
\qquad (\varepsilon\to0).
\end{equation}

Thus, although the soliton height vanishes, its
total mass remains finite. Defining the effective
soliton width as the ratio of its total mass to
its height, we obtain
\begin{equation}
W\equiv\frac{M}{H}
\simeq\frac{1}{\kappa(3\varepsilon^2)^{1/3}}.
\end{equation}

We therefore arrive at the scaling laws
\begin{equation}
\begin{aligned}
H&\propto\varepsilon^{2/3},\\
W&\propto\varepsilon^{-2/3},\\
M&\longrightarrow\frac{1}{\kappa}.
\end{aligned}
\end{equation}

\begin{figure*}[!htbp]
    \centering
    \includegraphics[width=\textwidth]{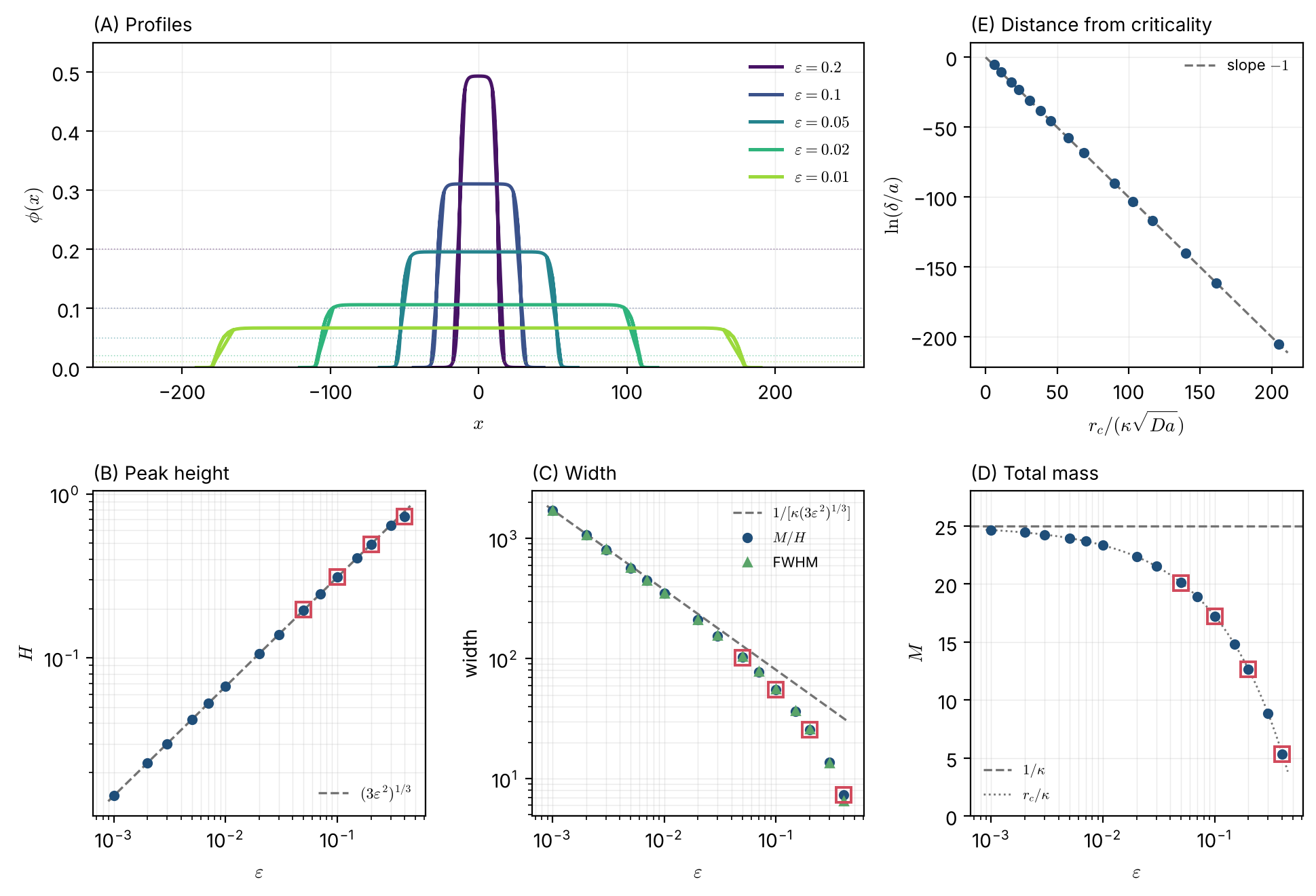}
    \caption{
\textbf{Numerical verification of soliton scaling laws.}
Results for fixed global competition strength
$\kappa=0.04$ and diffusion coefficient $D=1$.
\textbf{(A)} Stationary density profiles $\phi(x)$
on the broad soliton branch for decreasing values
of the reproduction cutoff $\varepsilon$.
As the cutoff decreases, the soliton becomes broader
and lower, developing a nearly flat plateau.
The horizontal dotted lines indicate the
corresponding cutoff values.
\textbf{(B)} Peak height $H$ as a function of
$\varepsilon$. The dashed line shows the
asymptotic prediction $H=(3\varepsilon^2)^{1/3}$.
\textbf{(C)} Effective width $W=M/H$ and full width
at half maximum (FWHM). Both approach the
parameter-free asymptotic prediction
$W=1/[\kappa(3\varepsilon^2)^{1/3}]$ (dashed line).
\textbf{(D)} Total mass $M$ as a function of
$\varepsilon$. The dashed horizontal line marks
the limiting value $1/\kappa$, while the dotted
curve shows the approximation $M\simeq r_c/\kappa$,
where $r_c=1-(3\varepsilon^2)^{1/3}$.
\textbf{(E)} Exponential approach to criticality.
Writing $a=(3\varepsilon^2)^{1/3}$ and
$\delta=r_c-r$, we plot $\ln(\delta/a)$ against
$r_c/(\kappa\sqrt{Da})$. The dashed line has
the theoretically predicted slope $-1$,
with a fitted additive constant.
Blue circles represent self-consistent stationary
solutions calculated from the full mechanical
energy integral. Red squares show the long-time
states obtained by direct numerical integration
of the time-dependent equation, confirming
convergence to the broad stationary branch.
The results in panel (E) were obtained independently
of the asymptotic approximation, by solving the
full self-consistency condition using
arbitrary-precision arithmetic with $\delta$
as the numerical parameter.
}
\label{fig:soliton_scaling}
\end{figure*}

Interestingly, the diffusion coefficient $D$ does
not appear in the leading-order expressions for
the soliton height and width. In this limit, the
height is determined by the reproduction cutoff,
the total mass by global competition, and the
width follows from their ratio.
These scaling laws are confirmed numerically in
Fig.~\ref{fig:soliton_scaling}(B)--(D).

We can further characterize how the broad soliton
approaches the critical point. Let us define
\begin{equation}
a=1-r_c=(3\varepsilon^2)^{1/3},
\qquad
\delta=r_c-r.
\label{eq:delta_def}
\end{equation}
As $\delta\to0^+$, the local maximum of the
effective potential approaches zero energy.
In the mechanical analogy, the particle spends
an increasingly long time near a quadratic maximum,
causing the soliton width to diverge logarithmically:
\begin{equation}
W\sim\sqrt{\frac{D}{a}}\,
\ln\left(\frac{10.34 a}{\delta}\right).
\label{eq:width_log}
\end{equation}
Since the profile becomes broad and nearly flat,
with a height approaching $a$, its total mass
also diverges logarithmically:
\begin{equation}
M\sim\sqrt{Da}\,
\ln\left(\frac{0.8187 a}{\delta}\right).
\label{eq:mass_log}
\end{equation}
However, the self-consistency condition requires
\begin{equation}
M=\frac{r}{\kappa}
\simeq\frac{1}{\kappa}.
\end{equation}

Consequently,
\begin{equation}
\ln\left(\frac{0.8187 a}{\delta}\right)
\sim\frac{1}{\kappa\sqrt{Da}}.
\label{eq:delta_exponential}
\end{equation}
Substituting this result into Eq.~(20)
gives $W\sim 1/(\kappa a)\propto\varepsilon^{-2/3}$,
in agreement with Eq.~(18). This result fixes the leading exponential
dependence of $\delta$ on $1/\kappa$,
$$\delta \simeq 0.8187 \  a\exp\{-1/(\kappa\sqrt{Da})\}.$$
Thus, self-consistency drives the stable soliton
exponentially close to the critical point as
$\varepsilon\to0$, as confirmed numerically in
Fig.~\ref{fig:soliton_scaling}(E). This proximity allows the
soliton to become extremely broad even as its
height vanishes.

For completeness, we briefly examine the asymptotic decay of the soliton tails. In the region where $\phi<\varepsilon$, reproduction is suppressed, and at sufficiently low densities the stationary equation reduces to
\begin{equation}
D\phi''-r\phi\simeq0.
\end{equation}
For $r>0$, the tails therefore decay exponentially,
\begin{equation}
\phi(x)\propto
\exp\left(-\sqrt{\frac{r}{D}}\,|x|\right),
\label{eq:exponential_tail}
\end{equation}
with a characteristic decay length
\begin{equation}
\ell_{\mathrm{tail}}=\sqrt{\frac{D}{r}}.
\end{equation}
Formally, when $r=0$, the local competition term must be retained, yielding an algebraic decay,
\begin{equation}
\phi(x)\sim\frac{6D}{x^2}.
\label{eq:algebraic_tail}
\end{equation}
However, this limit is not relevant to the broad self-consistent soliton considered here. As $\varepsilon\to0$ at fixed $\kappa$, we have $r\to r_c\to1$, so that $\ell_{\mathrm{tail}}\to\sqrt{D}$ remains finite. The divergence of the soliton width, $W\propto\varepsilon^{-2/3}$, therefore originates from the expansion of its nearly flat, populated interior rather than from the dilute tails.

\section{Demographic discreteness and deterministic establishment thresholds}
\label{sec:demographic}

We have so far represented the failure of small populations to establish themselves by a deterministic reproductive cutoff. This approximation has a natural demographic interpretation, but it is not equivalent to an explicit population of discrete individuals. A single individual arriving in an unoccupied niche can reproduce, although its descendants may subsequently become extinct through demographic fluctuations. In contrast, the deterministic cutoff suppresses reproduction altogether below a prescribed density. The question is therefore which predictions of the localized state survive when the cutoff is replaced by stochastic birth, death, and mutation. Previous work has shown that deterministic cutoffs can capture aspects of front propagation while missing spreading driven by stochastic cluster formation \citep{ganan2018front}. Our analytically characterized localized solution provides an opportunity to test this correspondence quantitatively.

We perform this comparison on a one-dimensional lattice of trait values, keeping the mutation and competition rules fixed while changing the description of the population. In the individual-based model, site $i$ contains an integer number $n_i$ of individuals, with density $\phi_i=n_i/N$ and total rescaled population $M=\sum_i\phi_i$. Each individual reproduces at rate one, hops to either neighboring site at rate $\mu$ per direction, and dies at a rate
\begin{equation}
\frac{n_i-1}{N}+\kappa M.
\label{eq:individual_death_rate}
\end{equation}
The first term represents competition with other individuals at the same site, and the second represents competition for shared global resources. There is no reproductive cutoff or Allee effect in this stochastic model: the per-capita reproduction rate is independent of local density. As a complementary approximation, we simulate demographic noise in a continuous-density lattice model using the scheme developed by \cite{weissmann2017simulation}, which builds on the split-step algorithm introduced by \cite{dornic2005integration}, with a leading-order noise variance of $2\phi_i/N$ per unit time. Finally, we evolve the deterministic lattice equation with a reproductive cutoff,
\begin{equation}
\dot\phi_i=\mu(\phi_{i+1}+\phi_{i-1}-2\phi_i)
+\phi_i\Theta(\phi_i-\varepsilon)
-\phi_i^2-\kappa M\phi_i.
\label{eq:deterministic_trait_lattice}
\end{equation}
The use of the same trait lattice separates the effects of discreteness in trait space from those of demographic fluctuations.

\begin{figure*}[!htbp]
\centering
\includegraphics[width=1\textwidth]{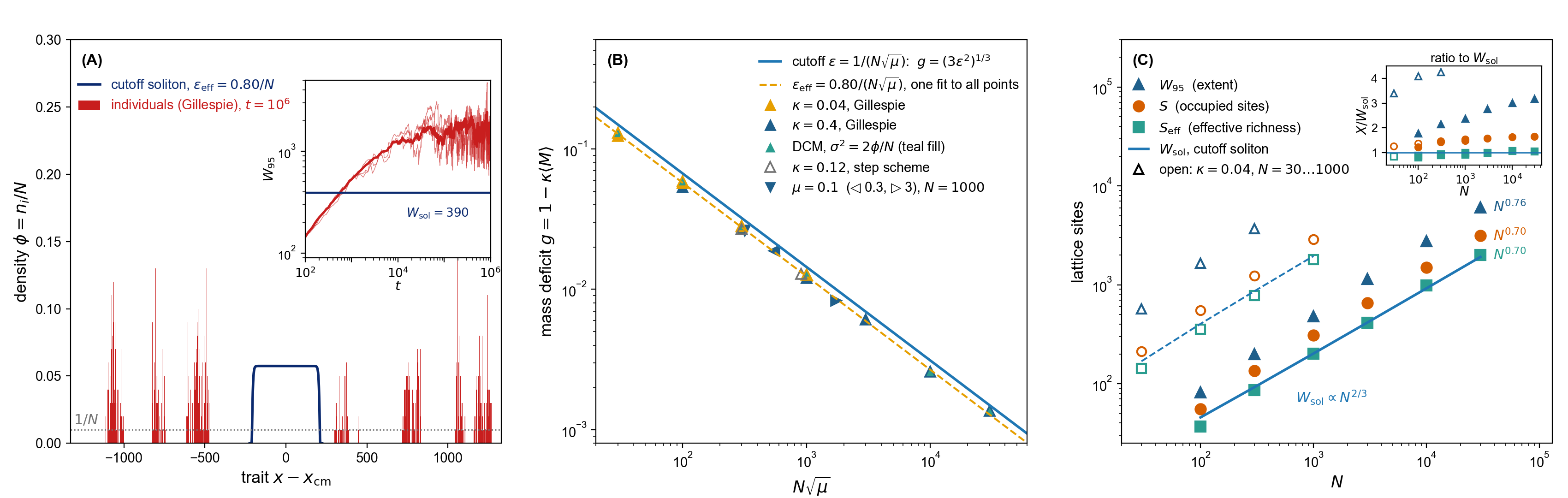}
\caption{\textbf{Comparison between the deterministic soliton
and the discrete-individual population.}
(A) Population density in trait space for $N=100$,
$\kappa=0.04$, and $\mu=1$. The dark-blue curve shows
the deterministic soliton with
$\varepsilon_{\mathrm{eff}}=0.80/N$; the red profile
is a Gillespie realization at $t=10^6$.
The dotted line marks the single-individual density $1/N$.
The inset shows the evolution of $W_{95}$ for three
realizations (thin lines), their eight-realization mean
(thick line), and the deterministic soliton width. (B) Stationary mass deficit $g=1-\kappa\langle M\rangle$
versus $N\sqrt{\mu}$. Orange and blue triangles show
Gillespie results for $\kappa=0.04$ and $0.4$,
respectively; teal-filled and open grey symbols show
split-step and discrete-time simulations.
Marker orientation distinguishes mutation rates.
For $\kappa=0.04$, averages are taken over the
stationary regime of 8 realizations (16 at $N=1000$).
The solid line is the analytical prediction
$g=(3\varepsilon^2)^{1/3}$ with
$\varepsilon=1/(N\sqrt{\mu})$; the dashed line uses
$\varepsilon_{\mathrm{eff}}=0.80/(N\sqrt{\mu})$,
with a single fitted coefficient. (C) Spatial extent $W_{95}$ (triangles), occupied
trait sites $S$ (circles), and effective species
richness $S_{\mathrm{eff}}$ (squares), for $\mu=1$.
Filled and open symbols correspond to $\kappa=0.4$
and $0.04$, respectively. Solid and dashed lines
show the corresponding deterministic soliton widths
$W_{\mathrm{sol}}$ with
$\varepsilon_{\mathrm{eff}}=0.80/N$.
The inset shows the ratios to $W_{\mathrm{sol}}$.
The indicated exponents are power-law fits over
the investigated range.}
\label{fig:demographic_comparison}
\end{figure*}

Demographic fluctuations are known to generate spontaneous spatial clustering even in the absence of attractive interactions between individuals \citep{meyer1996clustering,young2001brownian,houchmandzadeh2002clustering}. More directly relevant to evolutionary diversification, Rogers, McKane, and Rossberg demonstrated that demographic noise can produce clusters in trait space even when the corresponding deterministic dynamics does not predict pattern formation \citep{rogers2012demographic}. Figure~\ref{fig:demographic_comparison}(A) illustrates this effect in our model: the continuous deterministic soliton is replaced by an archipelago of spatially separated population clusters, each comprising individuals with similar traits. In an evolutionary interpretation, these clusters may be viewed as groups of closely related species, analogous to genera separated in trait space. Following an initial period of expansion, the stochastic population reaches a statistically stationary spatial extent, as shown in the inset, although its saturated width is substantially larger than that of the deterministic soliton. Individual clusters continue to fluctuate, migrate, appear, and disappear, so that the stochastic state is not a stationary density profile but a dynamically maintained, fragmented population. This pronounced difference in spatial organization raises a natural question: to what extent do the analytical predictions obtained for the deterministic soliton remain applicable to the stochastic population?

We begin with the total population mass. To compare the analytical
prediction for the deterministic soliton with the individual-based
model, we must identify the reproductive cutoff $\varepsilon$ with
an appropriate discreteness scale in the stochastic system. Since
a single individual at a lattice site corresponds to a density
$1/N$, this would appear to be the natural scale. However,
diffusion reduces local demographic fluctuations by mixing
individuals between neighboring sites. When distances are
measured in units of the diffusive length scale $\sqrt{\mu}$,
the demographic noise strength is controlled by the combination
$N\sqrt{\mu}$ rather than by $N$ alone. This suggests an
effective cutoff of the form
$\varepsilon_{\mathrm{eff}}=c/(N\sqrt{\mu})$, where the
numerical prefactor $c$ is not determined by this scaling
argument. For the deterministic soliton, we found that the
mass deficit obeys
$g=1-\kappa M\simeq(3\varepsilon^2)^{1/3}$.
Figure~\ref{fig:demographic_comparison}(B) shows that the
mean mass deficit in the stochastic model is well described
by the same expression, with
$\varepsilon_{\mathrm{eff}}\simeq0.8/(N\sqrt{\mu})$.
This agreement holds across different values of the
population-size parameter $N$, the mutation rate $\mu$,
and the global competition strength $\kappa$.

Next we examine whether a similar correspondence holds for the
extent and diversity of the stochastic population in trait space.
Figure~\ref{fig:demographic_comparison}(C) compares the width of
the deterministic soliton with three measures of the stochastic
population: the spatial extent $W_{95}$ containing 95\% of all
individuals, the number of occupied trait sites
$S=\sum_i\Theta(n_i)$, which may be interpreted as species
richness, and the effective species richness
$S_{\mathrm{eff}}=(\sum_i n_i)^2/\sum_i n_i^2$,
which accounts for differences in abundance among species. Despite the fragmented structure of
the stochastic population, both measures of richness exhibit
approximately the same $N^{2/3}$ scaling as the deterministic
soliton width. In particular, the ratio of occupied sites to
soliton width approaches a value of approximately 1.6, while
the ratio $S/S_{\mathrm{eff}}$ remains nearly constant.
The spatial extent $W_{95}$ also increases with $N$, but is
substantially larger and grows somewhat faster over the
investigated range, reflecting the empty gaps separating the
clusters. Thus, although the deterministic theory does not
reproduce the spatial organization of the stochastic population,
it approximately captures the scaling of the number of occupied
trait niches.

In summary, the comparison between the deterministic and
individual-based models shows that an effective reproductive
cutoff can approximately reproduce the scaling laws of total
population mass and species richness, even though it fails to
capture the spatial organization of the population. Whereas
the deterministic cutoff produces a continuous, localized
soliton, demographic fluctuations generate a dynamic
archipelago of separated clusters spanning a substantially
broader range of trait space. This distinction highlights
two different interpretations of the reproductive cutoff.
A genuine ecological establishment threshold, such as an
Allee effect, may arrest diversification through the
front-pinning mechanism analyzed here. Demographic
discreteness, in contrast, does not completely prevent
small populations from establishing, but instead gives
rise to stochastic dynamics in which clusters continually
emerge and disappear. Despite these differences, our
results suggest that a simple deterministic description
can correctly predict the scaling of global ecological
properties even when it fails to reproduce the spatial
organization of the community.

\section{Discussion}
\label{sec:discussion}

A central question in evolutionary ecology is what determines the extent of biological diversity when differentiation into new niches reduces competition among species but does not eliminate competition for shared resources. Our results suggest that resource limitation, even when it regulates the total population size, is not sufficient by itself to constrain the spread of a population in trait space: without an establishment threshold, the population occupies an ever-widening range of traits at ever lower density, whereas an arbitrarily small threshold allows a stable localized distribution. Thus, the extent of diversity depends not only on the availability of potential niches or the total amount of resources, but also on the ability of initially small populations to establish themselves in new niches. The analytical tractability of the model makes this dependence quantitative, $W\propto\varepsilon^{-2/3}$. Although the width of a trait distribution is not equivalent to species richness, it provides a quantitative measure of the extent of ecological differentiation and of the constraints acting upon it.

The localization mechanism identified here differs from the pattern-forming instabilities found in models with nonlocal competition, where trait-dependent interactions can destabilize homogeneous states and generate spatial patterns \citep{maruvka2006nonlocal,fuentes2003nonlocal,pigolotti2007species}. Our model has no such finite-wavelength instability. Instead, the establishment threshold, together with mortality imposed by global competition, creates bistability between empty and populated states, allowing localization through the arrest of invasion fronts. As the population expands, global competition increases and suppresses further invasion of unoccupied niches; as it contracts, competition weakens. In the small-cutoff regime, this feedback drives the system exponentially close to the Maxwell point, where invasion and retreat of an isolated front are balanced. The logarithmic divergence of the profile width near this point, combined with the global self-consistency condition, explains how a finite extent is selected without an externally imposed niche-scale length. Front arrest and localized colonies have previously been studied in systems with Allee effects and nonlocal interactions \citep{clerc2010analytical,keitt2001allee,maruvka2007nonlocal,kessler2024pulse}. The distinctive contribution of the present model is an analytically tractable description of how a local establishment threshold and global resource competition jointly select the extent of evolutionary diversification.

The establishment threshold admits two biological interpretations,
and distinguishing between them is essential. When it represents
an Allee effect, it describes a genuine ecological limitation on
the ability of small populations to establish themselves, and
the front-arrest mechanism analyzed here provides a possible
explanation for the limitation of diversification into new niches.
When the cutoff serves instead as a deterministic proxy for
demographic discreteness, a more complex picture emerges.
As established in previous studies, demographic fluctuations
can generate separated clusters in trait space even when the
corresponding deterministic dynamics does not predict such
patterns \citep{young2001brownian,houchmandzadeh2002clustering,rogers2012demographic}.
Nevertheless, our quantitative comparison shows that, despite
this fundamental difference in spatial organization, the
deterministic cutoff approximately reproduces the scaling of
the population mass deficit when identified with an effective
discreteness scale,
$\varepsilon_{\mathrm{eff}}\sim 1/(N\sqrt{\mu})$.
Moreover, both the number of occupied trait niches and the
effective species richness exhibit approximately the same
$N^{2/3}$ scaling as the deterministic soliton width.
Thus, the deterministic theory captures the scaling of global
community properties even though it fails to describe the
spatial organization of the population.

The difference between the two descriptions is particularly
evident in the role of rare events at the edges of the population.
In the deterministic cutoff model, reproduction is completely
suppressed below the threshold, preventing invasion fronts from
advancing beyond the populated region. In the stochastic model,
however, even a single individual arriving in an unoccupied
niche may reproduce and establish a new population. Such events,
together with local extinctions, generate a dynamic archipelago
of clusters that appear, disappear, and migrate through trait
space. Our simulations show that, following an initial period
of expansion, the spatial extent of this archipelago also reaches
statistical saturation, although its saturated width is
substantially larger than that of the deterministic soliton.
Furthermore, whereas the number of occupied trait niches
approximately follows the soliton scaling law, the spatial
extent grows somewhat faster over the investigated parameter
range, reflecting the empty gaps between clusters.
Thus, the mechanism determining the scale of ecological
diversity does not necessarily determine its spatial extent.
Deriving the observed scaling laws directly from the microscopic
stochastic dynamics, and understanding the factors controlling
the separation between clusters, remain open theoretical
questions.

Our results also have implications for eco-evolutionary models based on generalized Lotka--Volterra dynamics, in which new species inherit interaction coefficients similar to those of their ancestors \citep{ginzburg1988evolution,shtilerman2015emergence,farahpour2018tradeoff}. In such models, evolutionary differentiation may progressively weaken competition between species, potentially driving the community toward nearly noninteracting populations. Similarly, models with finite-range competition in niche space explicitly allow sufficiently dissimilar species to avoid competition altogether \citep{li2026clusters}. When such models are intended to describe communities constrained by shared limiting resources, this complete decoupling should be prevented. One simple implementation is to retain a positive baseline of competition, irrespective of the evolutionary distance between species. Our analysis shows, however, that this global competitive constraint limits total population size without necessarily limiting the extent of diversification in trait space. A mechanism suppressing the establishment of sufficiently small populations is also needed to obtain localization.

Finally, our model describes an idealized ecological system: a one-dimensional trait space, local competition between individuals with identical traits, and uniform global competition for shared resources. In natural communities, competition typically extends over a finite range of trait differences, while shared resources may themselves be heterogeneous or fluctuate in time. An important question is therefore how robust the front-arrest mechanism and the resulting scaling laws are to relaxing these assumptions. In particular, it would be interesting to consider smooth establishment thresholds and finite-range competition kernels, and to investigate under what conditions the clusters emerging from stochastic dynamics may develop into distinct evolutionary lineages. 

The role of the dimensionality of trait space is particularly interesting. In one dimension a mutant almost always lands in a niche that is already occupied, so that mutation acts as diffusion and neighboring populations continually rescue one another. In the opposite limit of a high-dimensional trait space a mutant never lands in an occupied niche; every mutation founds a new type, and the individual-based model of Sec. III reduces to an infinite-allele community in which each type is limited by its own density while all types share a common competitive load. This is the niche-stabilized neutral model solved by Pigolotti and Cencini~\cite{pigolotti2013species}, with their stabilization parameter corresponding to 1/N. In that limit there are no fronts and no soliton: each type fluctuates around its own carrying capacity until it is lost, and the mass deficit is set by the balance between origination and extinction of types. The scaling laws found here need not survive in that limit. 

More broadly, our results emphasize that the emergence of new ecological opportunities does not guarantee that they will be occupied. The fate of initially rare populations attempting to establish themselves in new niches may play a decisive role in determining the extent of biological diversity.

\clearpage
\bibliography{ref}

\clearpage
\appendix

\section*{Appendices}

\section{Absence of localization without a cutoff}
\label{app:no_cutoff}

Here we show that, in the absence of a reproduction
cutoff, global competition can limit the total
population but cannot prevent its spreading in
trait space. In particular, any initially localized
population must broaden indefinitely.

Setting $\varepsilon=0$, the evolution equation becomes
\begin{equation}
\partial_t\phi
=
D\partial_x^2\phi
+\phi-\phi^2-\kappa M\phi,
\qquad
M(t)=\int_{-\infty}^{\infty}\phi(x,t)\,dx.
\label{eq:appA_model}
\end{equation}
We consider nonnegative, nonzero initial conditions
with finite total mass and sufficient regularity and
decay for the following integrations to be valid.

To separate the total population from its spatial
distribution, we introduce the normalized density
\begin{equation}
p(x,t)=\frac{\phi(x,t)}{M(t)},
\qquad
\int_{-\infty}^{\infty}p(x,t)\,dx=1.
\label{eq:appA_p}
\end{equation}
Integrating Eq.~\eqref{eq:appA_model} over space gives
\begin{equation}
\frac{dM}{dt}
=
M-\kappa M^2-M^2Q,
\qquad
Q(t)=\int_{-\infty}^{\infty}p^2(x,t)\,dx.
\label{eq:appA_mass}
\end{equation}
Substituting $\phi=Mp$ into
Eq.~\eqref{eq:appA_model}, and using
Eq.~\eqref{eq:appA_mass}, we obtain
\begin{equation}
\partial_t p
=
D\partial_x^2p+Mp(Q-p).
\label{eq:appA_normalized}
\end{equation}
Notice that the global competition term
$-\kappa M\phi$ has disappeared explicitly from
the equation for the normalized distribution.
Global competition regulates the total population,
but does not directly favor one position in trait
space over another.

The quantity $Q=\int p^2\,dx$ provides a convenient
measure of localization. For a distribution that
is approximately uniform over a region of width
$W$, normalization gives $p\sim1/W$, and hence
$Q\sim1/W$. We may therefore define an effective
width
\begin{equation}
W_{\mathrm{eff}}(t)=\frac{1}{Q(t)}.
\label{eq:appA_width}
\end{equation}
To establish unlimited spreading, it is sufficient
to show that $Q(t)$ tends to zero.

Differentiating $Q$ and using
Eq.~\eqref{eq:appA_normalized}, we find
\begin{equation}
\frac{dQ}{dt}
=
-2D\int_{-\infty}^{\infty}(\partial_xp)^2\,dx
+
2M\left[
Q^2-\int_{-\infty}^{\infty}p^3\,dx
\right],
\label{eq:appA_Qdot}
\end{equation}
where we have integrated the diffusion term
by parts. By the Cauchy--Schwarz inequality,
\begin{equation}
Q^2
=
\left(\int p^2\,dx\right)^2
\leq
\left(\int p^3\,dx\right)
\left(\int p\,dx\right)
=
\int p^3\,dx.
\label{eq:appA_holder}
\end{equation}
Consequently, both terms on the right-hand
side of Eq.~\eqref{eq:appA_Qdot} are
nonpositive, and
\begin{equation}
\frac{dQ}{dt}
\leq
-2D\int_{-\infty}^{\infty}(\partial_xp)^2\,dx
\leq0.
\label{eq:appA_Qdecrease}
\end{equation}
Thus diffusion and local competition both act
to reduce localization.

To show that the spreading continues indefinitely,
rather than approaching a finite width, we use
the one-dimensional Nash inequality
\cite{nash1958continuity}. For a normalized
nonnegative density $p$, it states that
\begin{equation}
\int_{-\infty}^{\infty}(\partial_xp)^2\,dx
\geq C_N
\left(\int_{-\infty}^{\infty}p^2\,dx\right)^3
=C_NQ^3,
\label{eq:appA_nash}
\end{equation}
where $C_N>0$ is a universal constant.
Combining Eqs.~\eqref{eq:appA_Qdecrease}
and \eqref{eq:appA_nash} gives
\begin{equation}
\frac{dQ}{dt}\leq-2DC_NQ^3.
\label{eq:appA_Qbound}
\end{equation}
Integrating this differential inequality yields
\begin{equation}
Q(t)\leq
\left[Q(0)^{-2}+4DC_Nt\right]^{-1/2}.
\label{eq:appA_Qdecay}
\end{equation}
It follows that
\begin{equation}
W_{\mathrm{eff}}(t)
\geq
\left[Q(0)^{-2}+4DC_Nt\right]^{1/2}
\sim\sqrt{4DC_Nt}.
\label{eq:appA_spreading}
\end{equation}
Therefore, the effective width grows without bound,
at least as fast as $t^{1/2}$ at long times.
In particular, no nonzero localized stationary
solution can exist when $\varepsilon=0$.

Finally, we show that this unlimited spreading
is compatible with a finite total population.
Introducing $Y=1/M$, Eq.~\eqref{eq:appA_mass}
becomes
\begin{equation}
\frac{dY}{dt}=-Y+\kappa+Q(t).
\label{eq:appA_inverse_mass}
\end{equation}
Since Eq.~\eqref{eq:appA_Qdecay} establishes
$Q(t)\to0$, the solution of this linear equation
satisfies $Y(t)\to\kappa$. Hence
\begin{equation}
\boxed{
M(t)\longrightarrow\frac{1}{\kappa},
\qquad
W_{\mathrm{eff}}(t)\longrightarrow\infty.
}
\label{eq:appA_conclusion}
\end{equation}

Thus, without a cutoff, global competition
determines the asymptotic population mass but
does not confine its distribution in trait space.
The population becomes increasingly dilute as
it spreads over an ever-growing range of traits.
A reproduction cutoff is therefore essential
for the localized stationary solutions studied
in the main text.

\clearpage

\section{Stability of the localized solutions}
\label{app:stability}

We analyze the stability of localized stationary solutions of the full
nonlocal evolution equation introduced in the main text, which we
reproduce here for completeness:
\begin{equation}
\begin{gathered}
\frac{\partial\phi(x,t)}{\partial t}
=D\frac{\partial^2\phi(x,t)}{\partial x^2}
+\phi(x,t)\Theta\!\left(\phi(x,t)-\varepsilon\right)
-\phi^2(x,t)-\kappa M(t)\phi(x,t),\\
M(t)=\int_{-\infty}^{\infty}\phi(x,t)\,dx.
\end{gathered}
\label{eq:appB_full_model}
\end{equation}
As explained in the main text, stationary localized profiles can be
constructed by temporarily treating $r=\kappa M$ as an independent
parameter and solving
\begin{equation}
D\phi_r''+\phi_r\Theta(\phi_r-\varepsilon)
-\phi_r^2-r\phi_r=0
\label{eq:appB_stationary}
\end{equation}
for $\phi_r(x)$. The corresponding global competition strength is then
fixed by the self-consistency condition
\begin{equation}
\kappa(r)=\frac{r}{M(r)},
\qquad M(r)=\int_{-\infty}^{\infty}\phi_r(x)\,dx.
\label{eq:appB_branch}
\end{equation}
As shown in Fig.~2 of the main text, $\kappa(r)$ has a maximum
$\kappa_{\max}$. For any fixed $0<\kappa_0<\kappa_{\max}$,
the condition $\kappa(r)=\kappa_0$ admits two stationary localized
solutions, a narrow pulse and a broad pulse. We now examine their
dynamical stability.

For either stationary solution, let
$\phi(x,t)=\phi_r(x)+u(x,t)$, where $u$ is a small perturbation
and $\kappa_0=\kappa(r)$ is held fixed. Linearizing the full
evolution equation gives
\begin{equation}
\frac{\partial u}{\partial t}
=\mathcal{A}_r u
\equiv\mathcal{L}_r u
-\kappa_0\phi_r(x)\int_{-\infty}^{\infty}u(y,t)\,dy,
\label{eq:appB_linearized}
\end{equation}
where
\begin{equation}
\mathcal{L}_r
=D\partial_x^2+\left.\frac{df_\varepsilon}{d\phi}\right|_{\phi=\phi_r}-r,
\qquad
f_\varepsilon(\phi)=\phi\Theta(\phi-\varepsilon)-\phi^2.
\label{eq:appB_local_operator}
\end{equation}
For the sharp cutoff, $df_\varepsilon/d\phi$ is understood in the
distributional sense (or as the limit of a smooth cutoff).

To identify the zero mode associated with the fold, we differentiate
the stationary profile equation~\eqref{eq:appB_stationary}
with respect to the parameter $r$, obtaining
\begin{equation}
\mathcal{L}_r\,\partial_r\phi_r=\phi_r.
\label{eq:appB_branch_derivative}
\end{equation}
Applying the full linearized operator and using
Eq.~\eqref{eq:appB_branch_derivative}, we obtain
\begin{equation}
\begin{aligned}
\mathcal{A}_r\,\partial_r\phi_r
&=\phi_r-\kappa(r)\phi_r\int\partial_r\phi_r\,dx\\
&=\left[1-\kappa(r)\frac{dM}{dr}\right]\phi_r\\
&=M(r)\frac{d\kappa}{dr}\phi_r.
\end{aligned}
\label{eq:appB_fold_identity}
\end{equation}
Here we used $\int\partial_r\phi_r\,dx=dM/dr$ and
$M(r)\,d\kappa/dr=1-\kappa(r)\,dM/dr$, the latter following
directly from $\kappa(r)=r/M(r)$.
At the maximum of the self-consistency curve,
$d\kappa/dr=0$, and hence
\begin{equation}
\mathcal{A}_r\,\partial_r\phi_r=0.
\label{eq:appB_fold_zero}
\end{equation}
Thus the fold is accompanied by a neutral mode corresponding
to a change in the stationary pulse profile. This mode is distinct
from the neutral translation mode, which exists for every $r$.

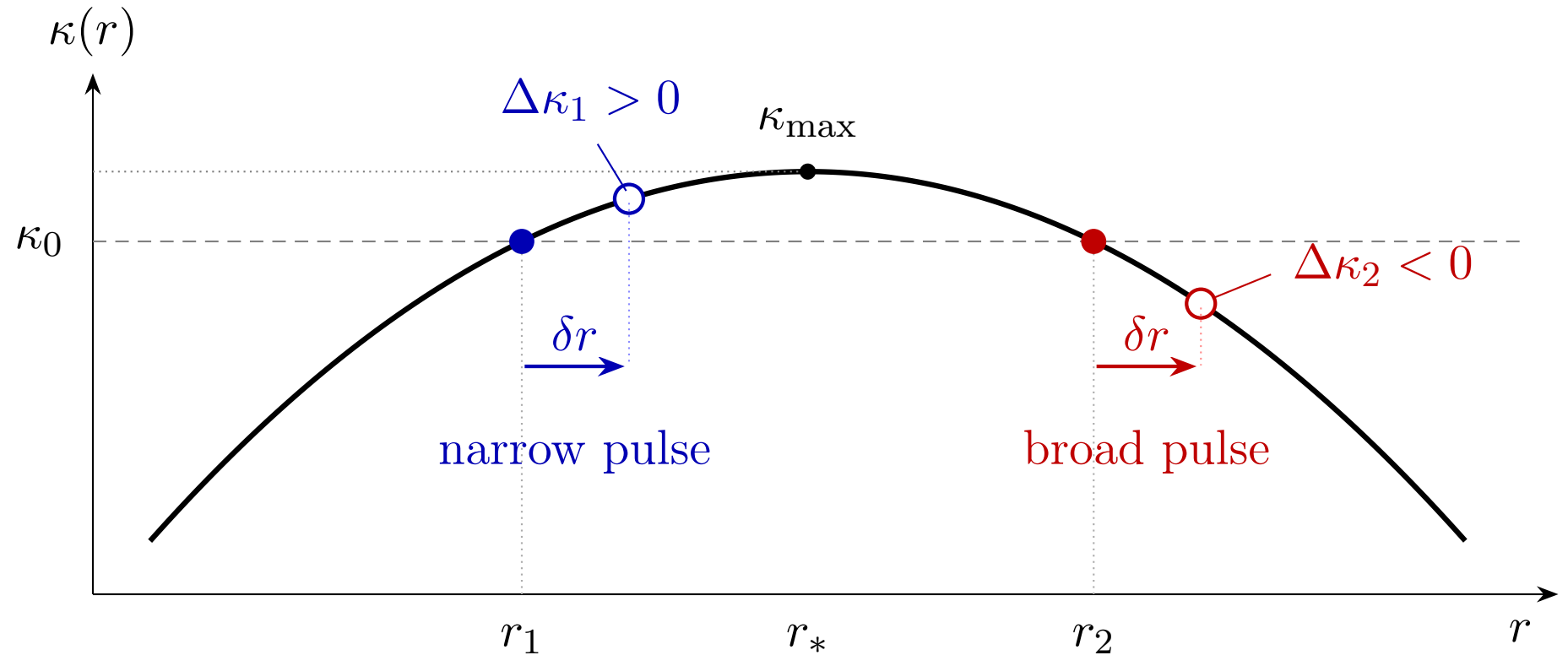
\begin{figure}[htbp]
\centering
\begin{tikzpicture}[x=1.12cm,y=0.72cm,>=Stealth,font=\small]
  \draw[->,thin] (0,0) -- (10.25,0) node[below left=2pt] {$r$};
  \draw[->,thin] (0,0) -- (0,5.67) node[above] {$\kappa(r)$};
  \draw[black,very thick,smooth,domain=0.40:9.60,samples=120]
    plot (\x,{4.6-0.19*(\x-5)^2});
  \draw[gray,densely dashed] (0,3.84) -- (10.02,3.84);
  \node[anchor=east] at (-0.08,3.84) {$\kappa_0$};
  \draw[gray,densely dotted] (0,4.6) -- (5,4.6);
  \fill (5,4.6) circle (1.8pt);
  \node[above=4pt] at (5,4.6) {$\kappa_{\max}$};
  \draw[gray!65,densely dotted] (3,0)--(3,3.79);
  \draw[gray!65,densely dotted] (7,0)--(7,3.79);
  \node[below=3pt] at (3,0) {$r_1$};
  \node[below=3pt] at (5,0) {$r_*$};
  \node[below=3pt] at (7,0) {$r_2$};
  \fill[blue!70!black] (3,3.84) circle (2.8pt);
  \fill[red!75!black] (7,3.84) circle (2.8pt);
  \fill[white] (3.75,4.303125) circle (3.2pt);
  \draw[blue!70!black,thick] (3.75,4.303125) circle (3.2pt);
  \fill[white] (7.75,3.163125) circle (3.2pt);
  \draw[red!75!black,thick] (7.75,3.163125) circle (3.2pt);
  \draw[blue!45,densely dotted] (3.75,4.26)--(3.75,2.49);
  \draw[red!45,densely dotted] (7.75,3.12)--(7.75,2.49);
  \draw[-{Stealth[length=2.1mm]},blue!70!black,thick]
    (3.02,2.48)--(3.72,2.48);
  \node[blue!70!black,fill=white,inner sep=1.4pt,above=2pt]
    at (3.37,2.48) {$\delta r$};
  \draw[-{Stealth[length=2.1mm]},red!75!black,thick]
    (7.02,2.48)--(7.72,2.48);
  \node[red!75!black,fill=white,inner sep=1.4pt,above=2pt]
    at (7.37,2.48) {$\delta r$};
  \node[blue!70!black,font=\small] at (3.37,1.55) {narrow pulse};
  \node[red!75!black,font=\small] at (7.37,1.55) {broad pulse};
  \node[blue!70!black,anchor=south] at (3.48,4.97)
    {$\Delta\kappa_1>0$};
  \draw[blue!70!black,thin] (3.53,4.90)--(3.73,4.39);
  \node[red!75!black,anchor=west] at (8.27,3.58)
    {$\Delta\kappa_2<0$};
  \draw[red!75!black,thin] (8.24,3.48)--(7.83,3.22);
\end{tikzpicture}
\caption{Schematic enlargement of the fold in Fig.~2 of the main text.
At $\kappa_0<\kappa_{\max}$, the two filled points denote the narrow
and broad stationary pulses. Rightward shifts $\delta r>0$ lead to
nearby auxiliary stationary profiles (open points), requiring
$\kappa_0+\Delta\kappa_i$, with $\Delta\kappa_1>0$ and
$\Delta\kappa_2<0$. In the argument, $\kappa_0$ is very close to
$\kappa_{\max}$; their separation and the shifts are exaggerated
for visibility.}
\label{fig:appB_fold_schematic}
\end{figure}

To examine which side of the fold is stable, we choose a fixed
$\kappa_0$ just below $\kappa_{\max}$. As illustrated in
Fig.~\ref{fig:appB_fold_schematic}, the line $\kappa=\kappa_0$
intersects the self-consistency curve at $r_1<r_*$ (the narrow pulse)
and $r_2>r_*$ (the broad pulse). To test the stability of each
stationary solution $\phi_{r_i}$, with $i=1,2$, at fixed $\kappa_0$,
we replace it by the nearby stationary profile $\phi_{r_i+\delta r}$,
where $\delta r>0$. This new profile corresponds to
$\kappa(r_i+\delta r)=\kappa_0+\Delta\kappa_i$.
We then use it as the initial condition for the dynamics at the
original competition strength $\kappa_0$.
On the rising branch $\Delta\kappa_1>0$, whereas on the falling
branch $\Delta\kappa_2<0$.

Write $\phi_i^+=\phi_{r_i+\delta r}$ and
$M_i^+=M(r_i+\delta r)$. The shifted profile is stationary at
$\kappa_0+\Delta\kappa_i$, and therefore satisfies
\begin{equation*}
0=D(\phi_i^+)''+f_\varepsilon(\phi_i^+)
-(\kappa_0+\Delta\kappa_i)M_i^+\phi_i^+.
\end{equation*}
The evolution of the \emph{same} profile at the original
competition strength $\kappa_0$ instead obeys, at $t=0$,
\begin{equation*}
\left.\frac{\partial\phi}{\partial t}\right|_{t=0}
=D(\phi_i^+)''+f_\varepsilon(\phi_i^+)
-\kappa_0 M_i^+\phi_i^+.
\end{equation*}
Subtracting the stationary equation cancels the diffusion and local
growth terms, leaving
\begin{equation*}
\left.\frac{\partial\phi}{\partial t}\right|_{t=0}
=\Delta\kappa_i\,M_i^+\phi_i^+.
\end{equation*}
Integrating over space, we obtain the exact initial mass response:
\begin{equation}
\left.\frac{dM}{dt}\right|_{t=0}
=\Delta\kappa_i\,[M(r_i+\delta r)]^2.
\label{eq:appB_mass_feedback}
\end{equation}

To interpret the sign of this response, note that
$\kappa(r)=r/M(r)$ implies
$\left.dM/dr\right|_{r_*}=M(r_*)/r_*>0$ at the fold.
Thus the positive shift $\delta r$ increases the initial mass
\emph{on both branches}. For the narrow pulse,
$\Delta\kappa_1>0$, so the mass initially grows still further;
for the broad pulse, $\Delta\kappa_2<0$, so it initially decreases.
Thus the initial mass response reinforces the perturbation for the
narrow pulse and opposes it for the broad pulse. Although this is not
a complete stability analysis, it provides a plausible physical
explanation for the instability of the narrow pulse and the stability
of the broad one, consistent with the numerical time evolution.
This picture is also supported by Appendix~A, which shows that without
a reproduction cutoff ($\varepsilon=0$) a localized population spreads
indefinitely. The resulting expectation that the stable localized
solution becomes arbitrarily broad as $\varepsilon\to0$ is consistent
with the behavior of the broad branch established in the main text.

\end{document}